# Pronounced grain boundary network evolution in nanocrystalline Cu subjected to large cyclic strains

David B. Bober [1,3*], Thomas LaGrange [2], Mukul Kumar [1], Timothy J. Rupert [3**]

[1] Lawrence Livermore National Laboratory, Livermore, CA 94550, United States
[2] École Polytechnique Fédérale de Lausanne, Interdisciplinary Centre for Electron Microscopy EPFL ENT-R CIME-GE MXC 134 (Bâtiment MXC) Station 12 CH-1015 Lausanne, Switzerland
[3] Department of Mechanical and Aerospace Engineering, University of California, Irvine, CA 92697, USA
 *E-mail: bober1@llnl.gov
**E-mail: trupert@uci.edu

## *Abstract*

The grain boundary network of nanocrystalline Cu foils was modified by the application of cyclic loadings and elevated temperatures. Broadly, the changes to the boundary network were directly correlated to the applied temperature and accumulated strain, including a 300% increase in the twin length fraction. By independently varying each treatment variable, a matrix of grain boundary statistics was built to check the plausibility of hypothesized mechanisms against their expected temperature and stress/strain dependences. These comparisons allow the field of candidate mechanisms to be significantly narrowed. Most importantly, the effect of temperature and strain on twin length fraction were found to be strongly synergistic, with the combined effect being ~150% that of the summed individual contributions. Looking beyond scalar metrics, an analysis of the grain boundary network showed that twin related domain formation favored larger sizes and repeated twin variant selection over the creation of many small domains with diverse variants.

*1. Introduction*

Grain boundary engineering of a nanocrystalline metal has been an attractive goal for many years [1, 2]. The motivation is provided by the extremely high density of grain boundaries, which makes their character especially important [3]. The most successful attempts to control nanocrystalline grain boundary character have relied on creating a high density of growth twins in low stacking fault energy metals produced by sputtering [4] or electrodeposition [5]. This parallels conventional grain boundary engineering (i.e., grain boundary manipulation in coarse-grained materials), where improvements in properties have also come from an increase in the twin fraction [6, 7]. Other nanocrystalline grain boundary engineering techniques have used either in situ deposition treatments or post-deposition annealing to increase the $\Sigma 3$ fraction [8, 9]. Improvements in ductility [5], fatigue life [10], corrosion resistance [11], and conductivity [5] have been reported. While these materials have the high twin fractions typical of conventional grain boundary engineered metals, they also have columnar grain structures that pose new challenges. LaGrange et al. [12] showed that such a grain boundary network topology leaves these materials vulnerable to thermal coarsening, despite the supposed stabilizing effect of the twins. Recently, a warm cyclic deformation process has been used to increase the length fraction of twins in nanocrystalline Ni by 48% [13]. This type of thermomechanical approach is appealing because it does not rely on the deposition of lamellar nanotwins, and could theoretically be applied to bulk nanocrystalline materials, similar to the practice of conventional grain boundary engineering.

In addition to any potential practical importance, this type of deformation-based technique presents a case where the unusual deformation mechanisms of nanocrystalline metals are important. It is likely that the plastic deformation of a nanocrystalline metal will directly alter the grain boundary character distribution, which is in sharp contrast to conventional grain boundary engineering. In the latter, the applied strain serves primarily to accumulate strain energy that is useful for driving subsequent recrystallization that reconfigures the boundary network. By itself, the deformation applied in conventional treatments has little effect on boundary network statistics, though dislocation storage can introduce local changes along a given boundary [14]. However, in a nanocrystalline metal, it has been hypothesized that even modest plastic deformation can alter boundary character rather dramatically. For example, Panzarino et al. showed that even small cyclic strains can cause nanocrystalline grains to rotate relative to their neighbors and create new twin boundaries [15]. Other proposed nanocrystalline deformation mechanisms could be associated with even more dramatic changes to the boundary network. For example, Froseth et al. reported on a profusion of stacking faults created via grain boundary nucleated dislocations that subsequently thicken into twins [16]. Alternatively, deformation induced de-twinning could dramatically reconfigure microstructure. Some nanocrystalline deformation phenomena, like stress-induced grain boundary migration [17], would lead to subtle changes in the boundary network that could nonetheless accumulate during cyclic loading [18].

To the extent that a nanocrystalline grain boundary network is influenced by plasticity, studying its evolution provides a means to probe that deformation mechanisms. The two-part goal of this work is to describe nanocrystalline grain boundary engineering in a way that both (1) guides processing and (2) provides insight about the relevant deformation physics. From a processing perspective, there are currently many open questions. For example, while increasing the number of stress cycles has been shown to increase the extent of microstructural changes, the effect of stress/strain amplitude is unclear [13, 18]. From a mechanistic perspective, there is a great deal to be learned from the analysis of the grain boundary network connectivity, with has so far only been described in nanocrystalline metals via triple junction statistics (see, e.g. [13]). Bober et al.

recently used analysis of the grain boundary network to reveal how the formation and growth of individual twin related domains proceeds via recrystallization and is responsible for the high twin fractions observed in conventional grain boundary engineered Ni [19].

In this paper, we report on the evolution of the grain boundary network in nanocrystalline Cu under warm cyclic strains. The range of strains and cycling temperatures expands on those employed in our previous work on nanocrystalline Ni [13], which allows the individual and combined effects of temperature and deformation to be more clearly distinguished. In addition, the procurement of larger orientation maps than were achieved in our previous work allows new types of information to be gathered about the topology of the grain boundary network. Larger maps provide the ability to reconstruct twin related domains and so measure their size and the frequency of each twinning variant. Together, these observations are used to identify the deformation mechanism most likely responsible for the alteration of the nanocrystalline grain boundary network.

## 2. *Materials and Methods*

Nanocrystalline Cu films were sputter deposited with a mean grain size of 15 ± 8 nm, as measured via transmission electron microscopy. This grain size is within the nanocrystalline regime, but still large enough for successful orientation mapping [20]. Warm cyclic strains were applied to the nanocrystalline films because elevated temperature and cyclic strain can combine to cause grain boundary network evolution. For elevated temperature tests, the samples were warmed with hot air to either 60 ˚C or 100 ˚C. These temperatures were selected because they produced little or no grain growth in the absence of strain, yet were sufficient to facilitate grain boundary evolution when applied in combination with a cyclic stress [13]. The applied strain rates were chosen so that the material was always held at temperature for 70 minutes. Polymer substrates, namely 22 µm thick cellulose acetate, were used to reduce strain localization in the Cu films during loading, with the goal being to enable larger strains than would be possible for unsupported metal films.

We hypothesize here that the application of larger strains is desirable because it will increase the extent of grain boundary mediated deformation and the accompanying boundary network evolution. The metal-on-polymer concept is adapted from the flexible electronics community, where this type of composite has been shown to allow much larger strains to be applied to metal films than would be possible if they were free standing [21]. The film/substrate composite can be modeled as a Voigt iso-strain composite [22]. Identical average strains in the Cu and cellulose acetate were verified by using a scanning electron microscope (Phenom Pro-X, Eindhoven, NL) to check for cracks or delamination, which were not detected in any of the experiments presented here. Strong adhesion is required and was achieved by preparing the cellulose acetate with a dehydration bake and a titanium layer with ~1 nm thickness. To produce the nanocrystalline Cu layer, a 5 cm diameter planar direct current magnetron system (LAB Line, Kurt J. Lesker, Jefferson Hills, PA) was operated at an argon pressure of 4 milliTorr and a power of 40 W, yielding a deposition rate of 0.057 nm·s$^{-1}$. The pre-deposition chamber pressure was approximately 10$^{-6}$ Torr and the final thickness of the Cu layer was 103 ± 9 nm.

A standard load frame (Instron 5848, Norwood, MA) was used for mechanical testing. The specimens were approximately 3 mm wide, with a gauge length of 27 mm. Our goal for the loading treatment was to impart the highest amount of plastic strain possible. To this end, the specimens were loaded with a profile that consisted of a superimposed ramp and triangle wave, as shown in Figures 1(a) and (b). The peak strain, measured by crosshead displacement, slowly rises from 2% to 5%, with a constant 2% strain amplitude. The cellulose acetate remains elastic to strains of >2%, far more than a Cu film can sustain before plastically deforming. This means that during loading the metal film is deformed past its elastic limit in tension and then plastically until a total 2% strain is reached. Upon unloading, the Cu film will experience an elastic unload, followed by plastic compression until the cellulose acetate is relaxed. Although the cellulose acetate is relatively compliant, it can compress the metal film due to its much greater thickness if the interface remains strong and local anelastic polymer relaxation is limited. This approximates a fully reversing load on the Cu, which is an unusual feature in a tension-tension test geometry. Figure 1(a) shows how the unloading step only released 98.5% of the tensile strain from the prior loading, leading to a slow increase in the peak displacement that helps compensate for creep in the cellulose acetate over time at the elevated temperatures. In contrast, a simple triangular displacement profile would have led to ever diminishing elastic strains. The advantage of the applied loading profile is that it maintains a high elastic strain amplitude in the cellulose acetate, which therefore keeps its ability to apply a compressive load to the metal film. Figures 1(c) and (d) show stress and strain data from the composite films for room temperature (20 °C) and 100 °C, respectively. The stress is given as a fraction of the peak stress because the actual load partitioning between the metal and polymer is not known. Attempts to deconvolve the stresses in each component with a simple composite model failed because the properties of the cellulose acetate vary under the influence of cyclic loading and temperature, and these variations in the substrate properties masked the relatively small effect of the Cu layer, which is ~220 times thinner than the cellulose acetate. Since solving this problem would require a complete description of the temperature-dependent mechanical behavior of the polymer substrate, which is beyond the scope of this work, we leave our data in terms of the fraction of peak stress. The tests were all displacement controlled, with one set of experiments using a 2% tensile strain, followed by a 1.97% unloading strain, repeated 100 times, at a strain rate of $10^{-3}$ s$^{-1}$. This will be referred to as the "2% strain cycling" condition. A second type of loading profile was also used, but with a 1% tensile strain and 0.985% unload, at a strain rate of $5\times10^{-4}$ s$^{-1}$. This will be called the "1% strain cycling" condition in the text that follows.

Crystal orientation maps were collected with transmission Kikuchi diffraction (TKD), which provides the nanometer resolution necessary for such small grains [20, 23]. This technique is well suited to samples whose thickness exceeds the grain size because the diffraction signal originates from only the bottom several nm of material [20, 23]. Improved microscope stability allowed the maps in this work to be much larger (~$10^6$ nm$^2$) than our previous study (~$10^4$ nm$^2$ [13]). Samples were prepared for TKD by dissolving the cellulose acetate with acetone, rinsing in methanol and then isopropyl alcohol, and affixing to TEM grids with adhesive. The free-standing foils were further thinned by ion milling at 2 KeV and 1 mA (Gatan model 691, Pleasanton CA). These relatively low power parameters were selected to avoid damaging or heating the samples. Once loaded into the scanning electron microscope (SEM) (FEI SCIOS, Hillsboro OR), an in-chamber plasma cleaner was used for 90 s to reduce organic contamination prior to scanning.

All orientation maps were collected with a 2 nm step size and post-processed with a standard dilation clean-up using the Orientation Imaging Microscopy software package (EDAX, Mahwah NJ). Where error bars are given, they represent the 95% confidence interval based on the population of reconstructed boundaries. These error bars do not capture the uncertainty due to the necessity of reconstructing a grain boundary network from imperfect orientation data, where there are inevitably some pixels that cannot be reliably indexed. These missing pixels are prevalent at grain boundaries in any orientation mapping technique and are especially so when the interaction volume is large relative to the grain size. The total uncertainty for any given measure would therefore be larger than the error bar by an unknown factor, based on the extent of dilation required to reconstruct the boundary network.

Transmission electron microscopy (TEM) was also used to collect images of the microstructure and perform compositional mapping. Bright field imaging was performed on an FEI CM300 and Titan 80-300. Electron Energy Loss Spectrometry (EELS) was performed on a JEOL JEM 2200FS operated at 200 kV accelerating voltage and equipped with a Schottky field-emission gun and in-column omega filter.

## 3. Results and Discussion

3.1 Grain Coarsening with Thermomechanical Treatment

Bright field diffraction contrast TEM images, such as those presented in Figure 2, show that the initial, as-deposited microstructure had a nanocrystalline grain size of 15 nm. High angle annular dark field imaging revealed that the grains were interspersed with even smaller particles of ~2 nm diameter, primarily located at triple junctions (Figure 3(a)). Energy dispersive spectroscopy showed that these particles had high O content (Figure 3(b)) and were likely a form of copper oxide, but also contained traces of Ar trapped during the sputtering process. Such oxide dispersion has been shown to help stabilize grain size via boundary pinning (see, e.g., [24]). Since the oxides were formed during deposition, as evidenced by their Ar content, it is reasonable to assume that they contributed to the development of such a fine grain size in otherwise pure Cu films.

The effect of thermomechanical processing on the grain size and morphology was assessed with TEM. Bright field images of each processed material are shown in Figure 2, arranged according to the applied cyclic strain and temperature, but all at the same magnification. For example, the three images in the top row of Figure 2 show materials that were all heated to 100 ˚C for 70 minutes. The difference between these materials is the extent of cyclic strain that they were subjected to while hot. The leftmost material was not strained, while the rightmost underwent 100 strain cycles with a cyclic strain of 2%. The middle image corresponds to an intermediate level of strain, with a cyclic strain of 1%. From these three data points, it is clear that the application of larger plastic strains led to greater amounts of grain growth. The same trend can be seen in the middle and bottom rows of Figure 2, which are for materials strained at 60 ˚C and 20 ˚C, respectively.

Regarding the role of temperature, the images in the left-hand column of Figure 2 show that heating the materials to 100 ˚C was sufficient to cause only minor grain growth in the absence of stress. Likewise, the right-hand column shows that the grain growth which occurred under the influence of cyclic stress was greater at higher temperatures. This raises the question of whether stress and temperature are independent drivers of grain growth, or if there is a synergistic relationship. In the case of independent phenomena, the principal of superposition would apply and the grain growth due to cyclic strain at elevated temperature would be equal to the sum of the grain growth under strain-free annealing and that caused by room temperature straining. Determining if this is the case will help to understand the mechanism behind the evolution of the grain boundary network in nanocrystalline metals during cyclic straining.

Orientation maps provide a convenient means of quantitatively analyzing grain size, while also providing a basis for the measurement of more sophisticated microstructural metrics. Figure 4 shows example orientation maps for the as-deposited material and the materials that were cyclically strained with the 2% condition at several temperatures. In the left column of Figure 4, each pixel color represents a measured orientation and no post-processing has yet been performed on these maps. The black areas are those locations where no orientation could be determined.

Even in these unprocessed maps, many of the same trends from the TEM analysis are also apparent. The first step in quantitatively analyzing these maps was to assign orientations to the black pixels. This was done with a dilation process commonly used in the EBSD literature [25]. When this was complete, any grain composed of fewer than 4 pixels was eliminated and the dilation repeated. This threshold was a compromise balancing the elimination of mis-indexed points against the retention of small grains. The grain reconstruction used a 2° threshold to separate grains. Grain size was found by first computing the mean grain area, which was then converted to an equivalent circle diameter. These grain size measurements, presented in Figure 5(a), quantify the trends that were qualitatively shown in the TEM images. This plot also shows that the effect of temperature and strain cycling are synergistic, not independent. The total increase in grain size under the simultaneous action of cyclic strain (2%) and elevated temperature (100 °C) is 83% more than the sum of the two individual processes. The synergy of strain and temperature can also be observed by noticing that the slope of the 2% strain cycling data is much steeper than the uncycled data. The same synergistic effect is observed for the case of cyclic strain at 60 °C. This observation allows classic capillary-driven grain growth to be discounted as the dominant mechanism behind the evolution of our films because stress does not enhance the rate of this form of grain growth [26]. Even though grain boundary curvature is responsible for the thermal instability of nanocrystalline metals, and it is evident in the strain-free evolution here, the observed synergistic effect of temperature and strain requires an alternate explanation. Pole figures are not shown because no obvious texture modification accompanied these other changes.

Another effect of warm cyclic strain was to moderately change the grain size distribution, as shown in Figure 5(b). In this cumulative distribution plot, the vertical axis displays the fraction of grains smaller than the size listed on the horizontal axis, with the grain size normalized as grain area divided by mean grain area. A perfectly monodisperse grain size, shown by a delta function on a histogram, would appear in this plot as a unit step function centered at 1. Figure 5(b) shows that strain cycling led to a reduced slope around the normalized mean (value of 1), corresponding to a broadened distribution. This can also be observed qualitatively in Figures 2 and 4.

3.2 Evolution of Grain Boundary Character

The same set of orientation maps used to calculate grain size can also be used to analyze the types of grain boundaries that make up the grain boundary network. Using the measured grain orientations, each boundary was classified according to the coincident site lattice (CSL) criterion [27]. The middle column of Figure 4 shows the reconstructed grain boundaries overlaid on the Kikuchi pattern quality. $\Sigma 3$ boundaries are shown in red, $\Sigma 9$ boundaries in orange, $\Sigma 5$, $\Sigma 7$ and $\Sigma 11$ boundaries in blue, and all others in black. The relative length and number fraction of various CSL types was then computed in order to quantify the grain boundary character distribution. The most obvious change was an increase in the fraction of twin boundaries, as presented in Figures 5(c) and (d). This finding is similar in concept to our previous observations in nanocrystalline Ni [13], although of a very different magnitude. In the present case, the change is roughly 300% of what was reported for Ni. This is most likely due to the larger strains applied here to the Cu, which are ~5 times larger than what could could be applied to the Ni. Of course, the effects of stacking fault energy and homologous temperature cannot be discounted, both of which act alongside the greater strain to favor more evolution in these Cu films. None of the non-$\Sigma 3^n$, low $\Sigma$ boundaries underwent significant changes. As Figure 5(c,d) shows, the number and length fraction of twin boundaries both increased with strain cycling. The trends observed for $\Sigma 3$ fraction are very similar to those seen for grain size. In both cases, the changes were correlated with cyclic strain and temperature. Like grain size, the increase in $\Sigma 3$ fraction is also found to be caused by a synergistic combination of elevated temperature and strain, with the combined effect being from 20% or 40% larger than the summed individual effects, as measured by number or length fraction, respectively. A unexpected feature of Figure 5(c,d) is the small apparent drop in the Σ3 fraction caused by annealing at 60 °C in the absence of strain. We are unable to provide an explanation for this observation.

This observed temperature dependence helps narrow the list of possible evolution mechanisms because it argues against deformation twinning. While a novel form of deformation twinning has been proposed for nanocrystalline metals [28], it shares the same temperature dependence as conventional deformation twinning (i.e., it becomes easier at lower temperatures). The apparent inactivity of this mechanism is also supported by the observation that twin fraction and grain size increase in concert. Were deformation twinning prevalent, then the grain size would be expected to drop as existing grains became subdivided.

The Σ3 number and length fraction metrics reported in Figure 5 convey subtly different information. An increase in number fraction indicates that either more twins were added to the microstructure or that other boundaries were preferentially removed during the grain growth. In addition to these possibilities, the length fraction is also sensitive to the relative length of Σ3s compared to other boundaries. These metrics can be combined to give a measure of the average twin length by simply dividing the length fraction by the number fraction, with the resulting data presented in Figure 6(a). The average twin length is found to be 70-170% larger than the average length of other boundaries. This metric increases with thermomechanical treatment, presumably because twins have lower energy than other boundary types [29].

Cyclic strain also caused changes in the Σ3 boundaries' deviation from the ideal CSL misorientation. This misorientation is the same one which is compared against the Brandon criterion to determine CSL categorization [30]. Even within the population of interfaces that are counted as Σ3 boundaries, there exists a distribution of deviations from the perfect twin. Figure 6(b) shows that the 2% cycling treatment at 100 ˚C caused the length-weighted mean deviation from a perfect twin misorientation to drop to 1.4˚, down from 2.3˚ in the as-deposited material. A similar change was observed in the number-weighted mean CSL deviation, but this data is not presented here for brevity. The decrease in deviation from perfect misorientation could be caused by the lengthening or creation of coherent twins, or alternately by the shortening/destruction of incoherent segments. In addition, this result could occur through a change in the misorientation of existing Σ3 boundaries. The distribution of deviations from the ideal CSL was also examined, but no clear trends were observed beyond the shifting mean. It is important to note that the CSL framework only deals with the misorientation of the boundary, while grain boundary normal does not factor into the description of the grain boundary character. This is especially important to note for twin boundaries, as coherent and incoherent twins have very different energies and behaviors (see, e.g., the work of LaGrange et al. [12]). However, reliable grain boundary normal information is not available here due to the fact that TKD gives a two-dimensional view of the three-dimensional grain structure, and even the trace is difficult to determine accurately at these nanometric length scales.

3.3 Evolution of Grain Boundary Network Topology

While the aggregate statistics presented above are evidence of significant changes in the grain boundary network by themselves, the longer-range characteristics of the network can also help explain the physical processes at work. To provide metrics for tracking such features, the concept of twin related domains (TRDs) can be useful [31, 32]. TRDs are collections of grains mutually connected by twin boundaries [33]. Most simply, TRDs can be quantified by their size, as measured by the number of twinned grains they contain. These clusters are shown in the right-most column of Figure 4, with each TRD assigned a distinguishing color. Figure 7 plots the root-mean-squared (RMS) TRD size, which weights each TRD by its number of members. The size of these domains increased in concert with the twin fraction, as temperature and cyclic strain were increased. Detailed analysis of TRDs in conventional grain boundary engineered metals showed a similar connection between TRD size and twin fraction [19]. The increase in RMS TRD size can be further broken down by looking at the distribution of TRD sizes, as shown in Figure 8. The fraction of 1 member TRDs, or grains without twin boundaries, decreases rapidly with strain cycling, which can either result from twin nucleation or the removal of untwinned grains. The fraction of TRDs with size 2 and 3, the next most populous types, are shown in Figures 8(b) and (c), respectively. Perhaps most surprising is that the fraction of grains in 3-member TRDs experienced a much faster increase than those in 2-member TRDS, where any change is small and roughly of the same scale as the error bars on the data. This could occur if 1 member TRDs were bisected by new twins (each bounded by two twin boundaries) and jumped directly to being 3 member TRDs.

This observation also provides a point of comparison for the expected behavior of rotation-induced twinning, which is another of the hypothesized deformation mechanisms in the literature [34]. In this theory, neighboring nanocrystalline grains can rotate into low energy, low $\Sigma$ configurations. It has been proposed that these rotations could cause low angle boundaries to disappear, facilitating grain growth by coalescence [15, 35] and also that twins may form when boundaries may rotate into a $\Sigma 3$ relationship [34]. Panzarino et al. [34] observed this effect in molecular dynamics simulations that were conceptually similar to our experiments, though with a smaller grain size of ~5 nm. There is a specific probability that two neighboring grains would have the appropriate orientations to rotate into a twin relation, though the odds cannot be calculated because the relevant misorientation space has not been fully described [34]. Assuming independent initial orientations, there is an exponentially smaller chance that a third grain could also rotate to join such a TRD. This is contradictory to our observations on TRD size and indicates that rotation-induced twinning was not the operative mechanism here. Of course, this statistical argument does not hold if another initial orientation correlation or long-range phenomenon dominate the process.

The internal structure of twin related domains can also be assessed for changes by observing the frequency of certain orientation correlations between second nearest neighbors within a TRD, termed the probability of repeated twinning [19, 36]. To understand this metric, consider three grains connected by two twin boundaries. If the two twin boundaries are repeats of the same twin variant, then two of the three orientations will have a $\Sigma 1$ relationship. If the twinning variants are not repeats, then a $\Sigma 9$ relationship must exist. A graph representation for this is shown in Figure 9(a), with grains represented by the nodes (circles) and twin boundaries by the edges (straight lines). In Figure 9(a), twin variants are distinguished by letters and each unique orientation assigned a number. This is a simple case of the more general theory of $\Sigma 3^n$ related domains described by Reed and Kumar [33]. The probability of repeated twinning is found by considering the second nearest neighbors within every TRD and dividing the number of $\Sigma 1$ relationships by the sum of the $\Sigma 1$ and $\Sigma 9$ relationships. For illustration, a 4-member TRD schematic with a probability of repeated twinning of 0.33 is shown in Figure 9(b). As a real-world example, nanotwinned Cu films such as those studied by Lu et al. [37] would have a probability of repeated twinning approaching 1 because nearly every second-neighbor within a columnar TRD has the same orientation. The probability of repeated twinning for all samples is plotted in Figure 9(c), with only 3-member TRDs counted. One observation from this figure is that the probability of repeated twinning appears to drop with mechanical cycling, serving as additional evidence that deformation twinning is not the primary driver behind the increase in Σ3 fraction. This data supports the prior observations of (1) a strong temperature dependence and (2) slight grain coarsening which also rules out deformation twinning. Unlike most of the other trends observed, there is no clear temperature dependence in Figure 9. Interestingly, all of the probabilities of repeated twinning (0.64-0.93) were much higher than seen in the 3-member TRDs of conventional, coarse-grained Cu which form during recrystallization 0.3) [19].

3.4 Discussion of Shear-coupled Grain Boundary Motion as a Possible Mechanism

Shear-coupled grain boundary motion is a remaining possible mechanism to explain the observed grain size and grain boundary character evolution. In this phenomenon, shear stress causes a combination of normal and perpendicular motion of a grain boundary [38-40]. Such a mechanism has been observed to occur in stressed bicrystals at high temperatures [40, 41], and has also been implicated in several prior studies of deformation-related nanocrystalline grain growth [18, 42-44]. Legros et al. [44] documented this mechanism in nanocrystalline Al, using an in situ TEM experiment to observe rapid grain boundary motion near an opening crack tip [44]. Rupert et al. [43] geometrically separated the areas of maximum stress and strain, finding that the maximum amount of grain growth occurred at the location of maximum stress, an observation that is consistent with a shear coupling mechanism. Interestingly, stress-driven grain boundary motion can reverse direction under continuous loading because of fluctuating grain boundary structure and bifurcated shear coupling factors [44, 45]. A related phenomenon appears to be at work during the fatigue loading of nanocrystalline specimens, where marked increases in grain size has been observed [18]. At the extreme, this coarsening can lead to very large grains and initiate failure [46]. Inhomogeneous dislocation storage may also become important once the grains become large enough [38].

The observed grain growth in this study depended on the accumulated strain. In a material with weak strain hardening like nanocrystalline Cu, high accumulated strain does not always translate directly into high stress. However, this observation can be reconciled by considering how the total plastic strain influences the chance of any individual grain boundary experiencing a high stress. At high macroscopic stress but low strain, some grains will remain at low stress because of polycrystalline inhomogeneity. As strain increases, plastic deformation will cause different grains to become highly stressed. The extremely small grain size means that the motion of even a single dislocation through the crystal can cause large local strains and changes in stress [47]. Consequently, large global strains on the sample increase the chance that any given grain boundary will, at least temporarily, experience a stress sufficient to drive significant boundary migration. A similar argument applies to cycle count because accumulated plastic strain will cause a different stress distribution during each loading and unloading sequence. This is an important point because cycle count has been observed to correlate with nanocrystalline grain growth [13, 18]. Interestingly, mechanically-induced grain growth can also reduce the flow stress, which in turn would reduce the driver for further grain growth. Such a mechanism may be responsible for a microstructural rate of change that diminishes with increasing accumulated strain. Such a trend is supported by Figures 5 and 6, which show the 1% cyclic strain case caused much more than half as much microstructural evolution as the 2% cyclic strain case, despite nominally accumulating half the plastic strain.

Temperature dependence is another point to be considered in evaluating the potential role of shear-coupled motion. Winning et al. [40] showed that the velocity of grain boundaries propelled by a shear stress can be increased by several orders of magnitude in the presence of a modest increase in temperature, because shear-coupled grain boundary motion requires dislocation climb and vacancy diffusion [40]. While the exact change in grain boundary mobility is dependent on local atomic structure and therefore grain boundary character, an Arrhenius-type relation generally applies [41]. For example, the measured activation enthalpies in Al bicrystals are all on the order of 1 eV [41], denoting a strong temperature dependence. This indicates that shear-coupled grain boundary motion should generally lead to a positive correlation between temperature and grain growth. An exception has been observed in the case of very pure nanocrystalline Cu, where indentation-induced grain growth was greater at cryogenic temperature than at ambient [42]. However, the materials in the present study are relatively impure, containing nanoscale second phase pinning particles which will increase the role of thermal activation [48, 49]. Unpinning from these particles may also explain the broadening of the grain size distribution, with grains whose boundaries become unpinned being more susceptible to further growth.

Shear-coupled grain boundary motion also provides an explanation for the increase in twinning under warm cyclic strain conditions. The propensity of a moving boundary to produce twinning has been well established, with the mechanism thought to be either growth accidents [50, 51] or stacking fault packets [52]. While these classic results are based on boundaries driven by plastic strain energy or capillary forces, we expect that similar phenomena can occur for stress-driven boundary motion in nanocrystalline metals. An et al. [17] presented evidence for the formation of twins along stress driven boundaries, describing it as "mechanically-driven annealing twinning." We suggest the more descriptive phrase 'shear-coupled boundary migration twinning.'

Although less established in the literature than the capillary driven analog, such a mechanism is consistent with the twinning behavior we have observed. An increase in either temperature or strain would cause an increase in shear-coupled boundary motion, with a combined synergistic effect. Such increased boundary motion provides greater opportunity for twinning and agrees with all of the evolution trends observed in this study. On the other hand, many other potential mechanisms fail. Assuming the temperature dependence of the twinning probability per boundary migration distance is similar to the case of capillary driven boundaries, any effect over the temperature range studied would be negligible for this mechanism [53]. The stacking fault packet model would explain the tendency for 3-member TRDs to be formed, as the new twins would bisect existing grains. However, this would be at odds with the tendency for the probability of repeated twinning to drop as the number of twins increased. The growth accident model more readily explains the observed the drop in the probability of repeated twinning because the twin variants would be selected by the local boundary curvature, which is relatively unconstrained. From the perspective of growth accident formed twins, the high fraction of 3-member TRDs implies a boundary is likely to undergo multiple twinning once unpinned. However, given that there are large error bars on Figure 9, this inference should be regarded as preliminary.

In addition to the nucleation of twins along migrating boundaries, shear-coupled motion could also act to increase the twin content by lengthening pre-existing twins. These twins could be dragged by connecting boundaries that are themselves driven by the shear stress. The new area created along the twin boundary requires less energy than had it been a random high angle boundary, so extending the twin is the more energetically favorable event. A related mechanism through which shear-coupled motion may act to increase the twin fraction is the preferential removal of less stable boundaries [54]. A more mobile grain boundary (as compared to the stable twin) may be swept across a grain and into another boundary, removing itself from the microstructure. A similar type of preferential survival is thought to increase twin fraction during conventional grain growth [55, 56]. Both mechanisms would be expected to show a synergistic effect with temperature. Taking these observations together, shear-coupled boundary migration twinning does not suffer the inconsistencies that were noted for the other hypotheses available in the literature. Therefore, we consider it the most probable explanation for the grain size and grain boundary network evolution observed here.

## *Conclusions*

The magnitude of grain growth in cyclically strained nanocrystalline Cu was shown to depend strongly on cyclic strain amplitude and temperature, with a synergistic effect when applied together. A similar increasing trend was observed for the number and length fractions of $\Sigma 3$ boundaries, which was accompanied by an increase in the relative length of $\Sigma 3$s and a decrease in their deviation from the perfect CSL configuration. The TRDs that formed favored 3-member configurations over 2-member ones, as evidenced by the large change in the former and the negligible change in the latter, with a probability of repeated twinning that was high in all samples (>64%) but decreased with thermomechanical treatment. The observation that higher strains cause more evolution of the grain boundary network complements prior work showing that the number of stress cycles increases the extent of grain growth and twinning. Of the various possible mechanisms available in the existing literature, mechanically-driven grain boundary motion with coupled shear and normal movement provides the most complete explanation for the observations presented here. The results shown here are relevant to nanocrystalline metals processing, understanding possible changes during the service life of advanced nanostructured materials, and improving future means for nanocrystalline grain boundary engineering.

**Acknowledgments**

This work supported by the National Science Foundation through a CAREER Award No. DMR-1255305. The research was partly performed under the auspices of the U. S. Department of Energy by Lawrence Livermore National Laboratory under Contract DE-AC52-07NA27344. D.B.B. and M.K. were supported by the U.S. Department of Energy (DOE), Office of Basic Energy Sciences, Division of Materials Science and Engineering under FWP# SCW0939. D.B.B. also acknowledges the support of the Livermore Graduate Scholar Program at Lawrence Livermore National Laboratory.

**REFERENCES:**

[1] Watanabe T, Tsurekawa S. Toughening of brittle materials by grain boundary engineering, Mater Sci Eng A 387–389 (2004) 447-455.

[2] Watanabe T. Grain boundary engineering: historical perspective and future prospects, J Mater Sci 46 (2011) 4095-4115.

[3] Gleiter H. Nanocrystalline materials, Prog Mater Sci 33 (1989) 223-315.

[4] Zhang X, Anderoglu O, Hoagland RG, Misra A. Nanoscale growth twins in sputtered metal films, JOM 60 (2008) 75-78.

[5] Lu L, Shen YF, Chen XH, Qian LH, Lu K. Ultrahigh strength and high electrical conductivity in copper, Science 304 (2004) 422-426.

[6] Lin P, Palumbo G, Erb U, Aust KT. Influence of grain boundary character distribution on sensitization and intergranular corrosion of alloy 600, Scripta Metall Mater 33 (1995) 1387-1392.

[7] Randle V. Grain boundary engineering: an overview after 25 years, Mater Sci Tech 26 (2010) 253-261.

[8] Kobayashi S, Tsurekawa S, Watanabe T. A new approach to grain boundary engineering for nanocrystalline materials, Beilstein J. Nanotechnol. 7 (2016) 1829-1849.

[9] Ma H, La Mattina F, Shorubalko I, Spolenak R, Seita M. Engineering the grain boundary network of thin films via ion-irradiation: Towards improved electromigration resistance, Acta Materialia 123 (2017) 272-284.

[10] Shute CJ, Myers BD, Xie S, Li SY, Barbee TW, Hodge AM, Weertman JR. Detwinning, damage and crack initiation during cyclic loading of Cu samples containing aligned nanotwins, Acta Materialia 59 (2011) 4569-4577.

[11] Zhao Y, Cheng IC, Kassner ME, Hodge AM. The effect of nanotwins on the corrosion behavior of copper, Acta Materialia 67 (2014) 181-188.

[12] LaGrange T, Reed BW, Wall M, Mason J, Barbee T, Kumar M. Topological view of the thermal stability of nanotwinned copper, Appl Phys Lett 102 (2013).

[13] Bober DB, Kumar M, Rupert TJ. Nanocrystalline grain boundary engineering: Increasing Σ3 boundary fraction in pure Ni with thermomechanical treatments, Acta Materialia 86 (2015) 43-54.

[14] Kumar M, Schwartz AJ, King WE. Microstructural evolution during grain boundary engineering of low to medium stacking fault energy fcc materials, Acta Materialia 50 (2002) 2599-2612.

[15] Panzarino JF, Ramos JJ, Rupert TJ. Quantitative tracking of grain structure evolution in a nanocrystalline metal during cyclic loading, Model Sim Mater Sci Eng 23 (2015) 025005.

[16] Frøseth AG, Derlet PM, Swygenhoven HV. Twinning in nanocrystalline fcc metals, Adv Eng Mater 7 (2005) 16-20.

[17] An XH, Lin QY, Wu SD, Zhang ZF. Mechanically driven annealing twinning induced by cyclic deformation in nanocrystalline Cu, Scripta Materialia 68 (2013) 988-991.

[18] Padilla HA, Boyce BL. A review of fatigue behavior in nanocrystalline metals, Exp Mech 50 (2010) 5-23.

[19] Bober DB, Lind J, Mulay RP, Rupert TJ, Kumar M. The formation and characterization of large twin related domains, Acta Materialia 129 (2017) 500-509.

[20] Trimby PW. Orientation mapping of nanostructured materials using transmission Kikuchi diffraction in the scanning electron microscope, Ultramicroscopy 120 (2012) 16-24.

[21]     Lu N, Wang X, Suo Z, Vlassak J. Metal films on polymer substrates stretched beyond 50%, Appl Phys Lett 91 (2007) 221909.

[22]     Li T, Huang ZY, Xi ZC, Lacour SP, Wagner S, Suo Z. Delocalizing strain in a thin metal film on a polymer substrate, Mech Mater 37 (2005) 261-273.

[23]     Keller RR, Geiss RH. Transmission EBSD from 10 nm domains in a scanning electron microscope, J Microsc 245 (2012) 245-251.

[24]     Koch CC, Scattergood RO, Saber M, Kotan H. High temperature stabilization of nanocrystalline grain size: Thermodynamic versus kinetic strategies, J Mater Res 28 (2013) 1785-1791.

[25]     Humphreys FJ. Grain and subgrain characterisation by electron backscatter diffraction, J Mater Sci 36 (2001) 3833-3854.

[26]     Hillert M. On the theory of normal and abnormal grain growth, Acta Metall Mater 13 (1965) 227-238.

[27]     Grimmer H, Bollmann W, Warrington DH. Coincidence-site lattices and complete pattern-shift lattices in cubic-crystals, Acta Crystallogr A A 30 (1974) 197-207.

[28]     Zhu YT, Liao XZ, Wu XL. Deformation twinning in nanocrystalline materials, Prog Mater Sci 57 (2012) 1-62.

[29]     Schuh CA, Kumar M, King WE. Universal features of grain boundary networks in fcc materials, J Mater Sci 40 (2005) 847-852.

[30]     Brandon DG. Structure of high-angle grain boundaries, Acta Metall Mater 14 (1966) 1479-1484.

[31]     Lind J, Li SF, Kumar M. Twin related domains in 3D microstructures of conventionally processed and grain boundary engineered materials, Acta Materialia 114 (2016) 43-53.

[32]     Reed BW, Kumar M, Minich RW, Rudd RE. Fracture roughness scaling and its correlation with grain boundary network structure, Acta Materialia 56 (2008) 3278-3289.

[33]     Reed BW, Kumar M. Mathematical methods for analyzing highly-twinned grain boundary networks, Scripta Materialia 54 (2006) 1029-1033.

[34]     Panzarino JF, Pan Z, Rupert TJ. Plasticity-induced restructuring of a nanocrystalline grain boundary network, Acta Materialia 120 (2016) 1-13.

[35]     Moldovan D, Wolf D, Phillpot S, Haslam A. Role of grain rotation during grain growth in a columnar microstructure by mesoscale simulation, Acta Materialia 50 (2002) 3397-3414.

[36]     Lin B. Investigating annealing twin formation mechanisms in face-centered cubic nickel, Ph.D. Thesis, Carnegie Mellon University (2015)

[37]     Lu L, Chen X, Huang X, Lu K. Revealing the maximum strength in nanotwinned copper, Science 323 (2009) 607-610.

[38]     Gianola DS, Van Petegem S, Legros M, Brandstetter S, Van Swygenhoven H, Hemker KJ. Stress-assisted discontinuous grain growth and its effect on the deformation behavior of nanocrystalline aluminum thin films, Acta Materialia 54 (2006) 2253-2263.

[39]     Cahn JW, Taylor JE. A unified approach to motion of grain boundaries, relative tangential translation along grain boundaries, and grain rotation, Acta Materialia 52 (2004) 4887-4898.

[40]     Winning M, Gottstein G, Shvindlerman LS. Stress induced grain boundary motion, Acta Materialia 49 (2001) 211-219.

[41]     Gorkaya T, Molodov DA, Gottstein G. Stress-driven migration of symmetrical $<1\,0\,0>$ tilt grain boundaries in Al bicrystals, Acta Materialia 57 (2009) 5396-5405.

[42]     Zhang K, Weertman JR, Eastman JA. Rapid stress-driven grain coarsening in nanocrystalline Cu at ambient and cryogenic temperatures, Appl Phys Lett 87 (2005).

[43]     Rupert TJ, Gianola DS, Gan Y, Hemker KJ. Experimental observations of stress-driven grain boundary migration, Science 326 (2009) 1686-1690.

[44]     Legros M, Gianola DS, Hemker KJ. In situ TEM observations of fast grain-boundary motion in stressed nanocrystalline aluminum films, Acta Materialia 56 (2008) 3380-3393.

[45]     Thomas SL, Chen K, Han J, Purohit PK, Srolovitz DJ. Reconciling grain growth and shear-coupled grain boundary migration, Nature Communications 8 (2017) 1764.

[46]     Furnish TA, Mehta A, Van Campen D, Bufford DC, Hattar K, Boyce BL. The onset and evolution of fatigue-induced abnormal grain growth in nanocrystalline Ni-Fe, J Mater Sci 52 (2017) 46-59.

[47]     Bitzek E, Derlet PM, Anderson PM, Van Swygenhoven H. The stress–strain response of nanocrystalline metals: a statistical analysis of atomistic simulations, Acta Materialia 56 (2008) 4846-4857.

[48]     Lücke K, Stüwe HP. On the theory of impurity controlled grain boundary motion, Acta Metall Mater 19 (1971) 1087-1099.

[49]     Gore MJ, Grujicic M, Olson GB, Cohen M. Thermally activated grain-boundary unpinning, Acta Metall Mater 37 (1989) 2849-2854.

[50]     Carpenter HCH, Tamura S. The formation of twinned metallic crystals, Proc R Soc A 113 (1926) 161-U131.

[51]     Gleiter H. Formation of annealing twins, Acta Metall Mater 17 (1969) 1421-&.

[52]     Dash S, Brown N. An investigation of the origin and growth of annealing twins, Acta Metall Mater 11 (1963) 1067-1075.

[53]     Pande CS, Imam MA, Rath BB. Study of annealing twins in fcc metals and alloys,, Metall Trans A 21 (1990) 2891-2896.

[54]     Frøseth AG, Van Swygenhoven H, Derlet PM. Developing realistic grain boundary networks for use in molecular dynamics simulations, Acta Materialia 53 (2005) 4847-4856.

[55]     Form W, Gindraux G, Mlyncar V. Density of annealing twins, Metal Sci 14 (1980) 16-20.

[56]     Gindraux G, Form W. New concepts of annealing-twin formation in face-centered cubic metals, J Inst Metals 101 (1973) 85-93.

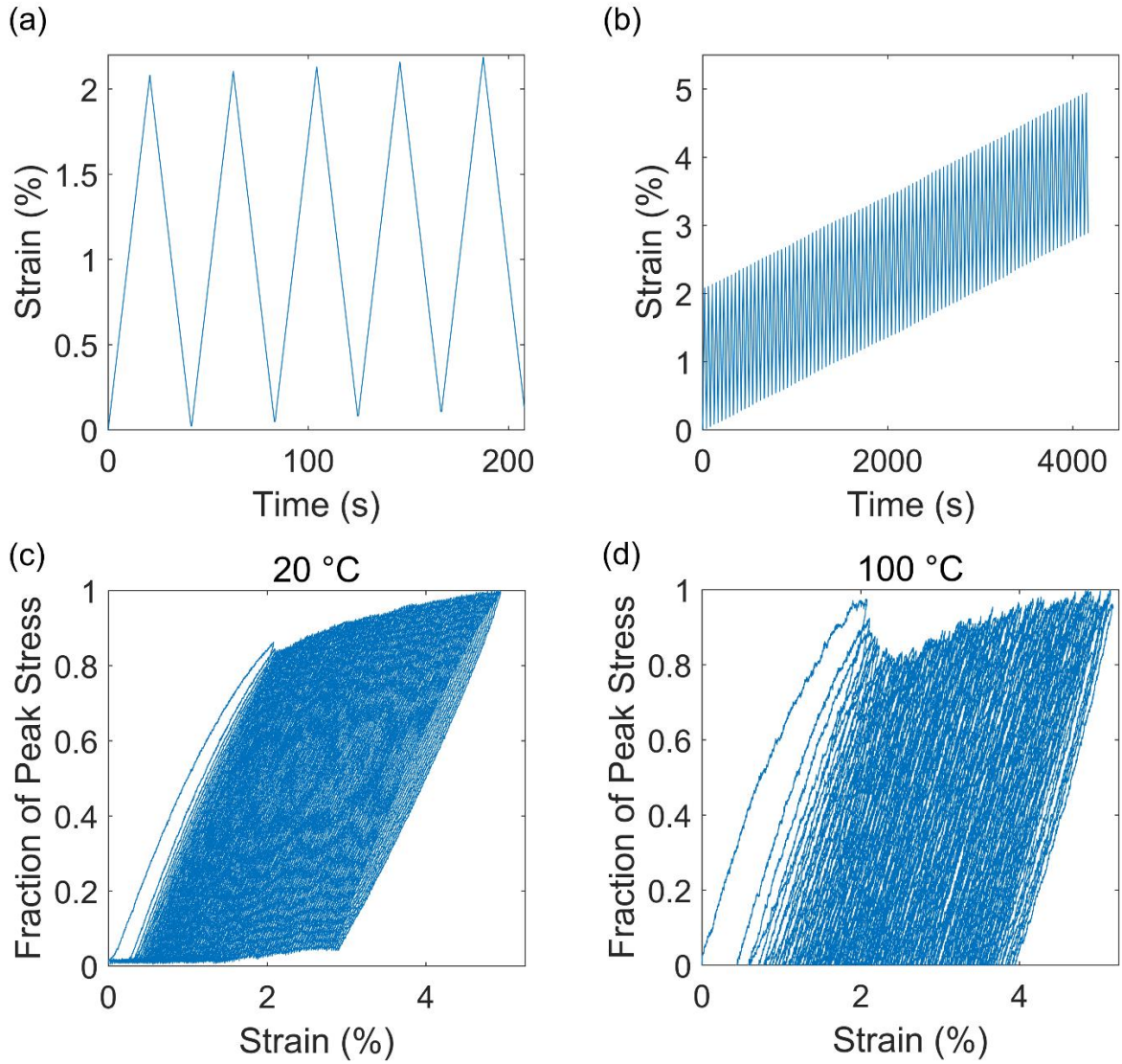

**Figure 1:** The 2% strain cycling applied to the cellulose acetate/copper composite, with the first 5 strain cycles shown in (a) and 100 strain cycles shown in (b). The mechanical response is shown for (c) 20 °C and (d) 100 °C testing temperatures.

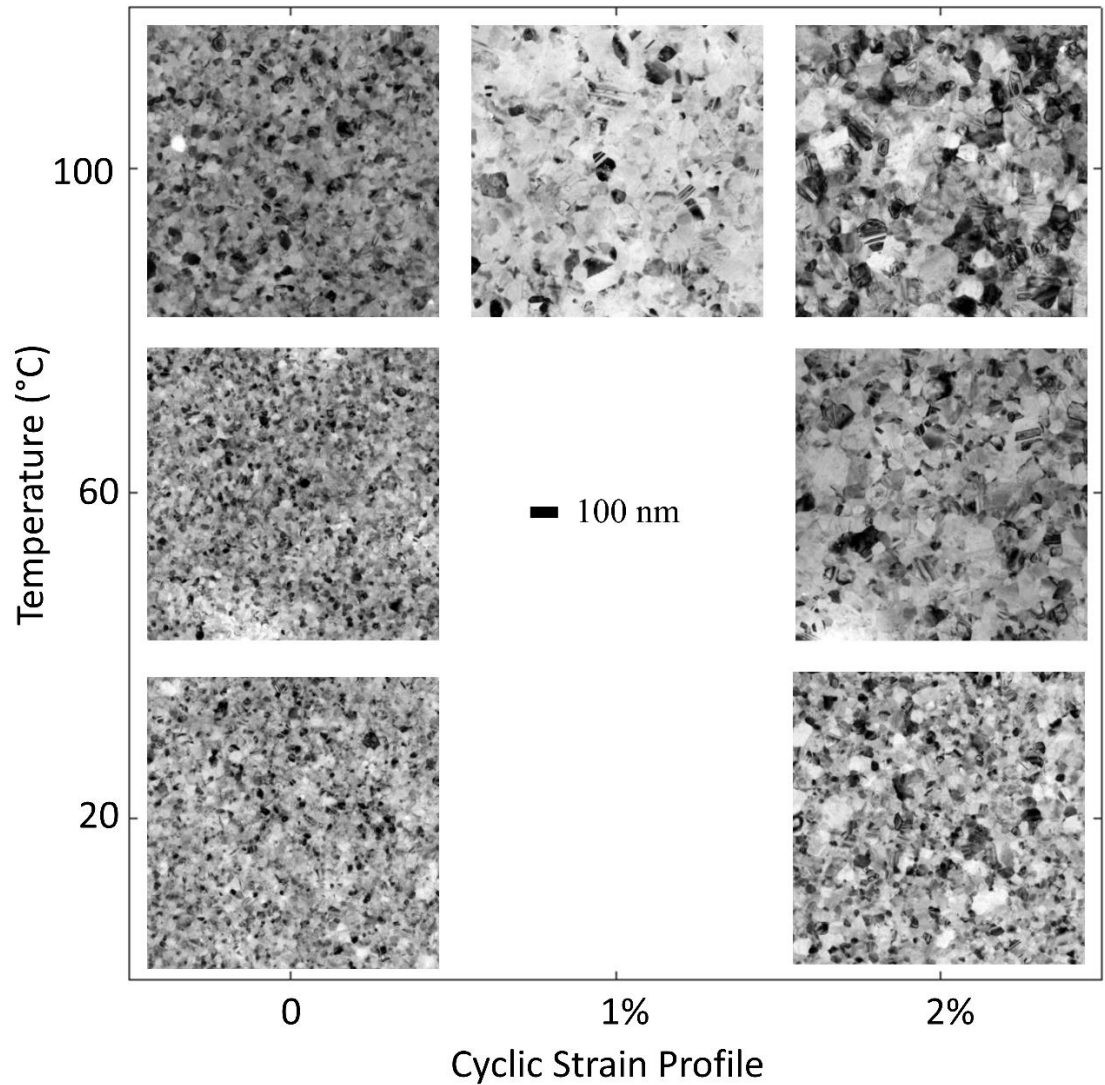

**Figure 2: Bright field TEM images of each Cu sample, with each image's location denoting the thermomechanical process applied. The testing temperature is labeled on the vertical axis and magnitude of cyclic strain on the horizontal axis.**

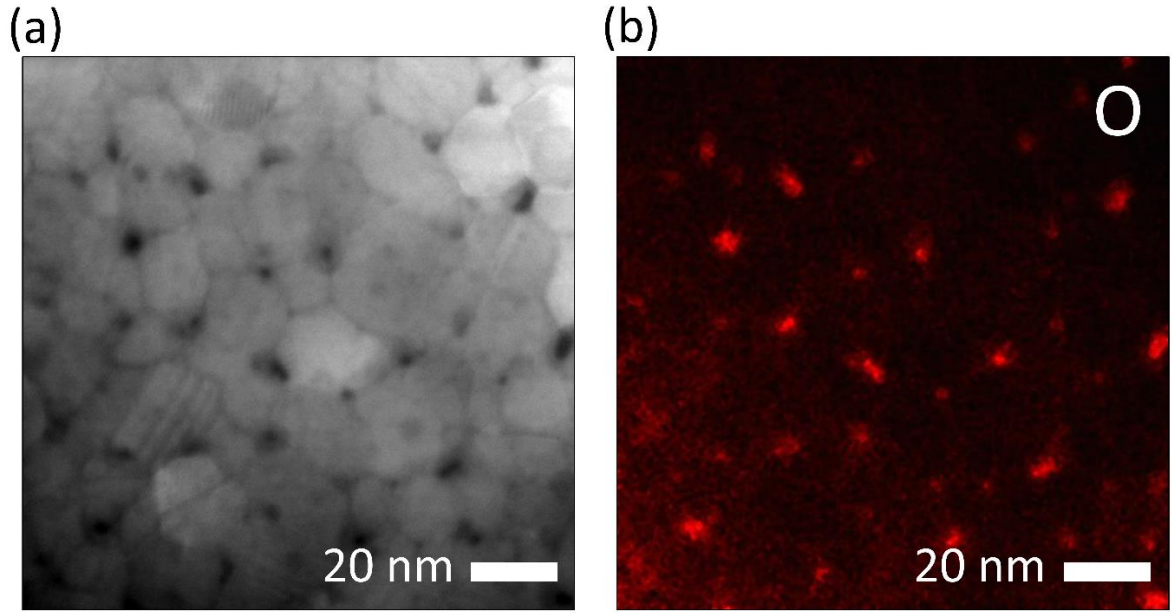

**Figure 3: (a) High-angle annular dark field image of the as-deposited material, showing particles at the grain boundaries. (b) Energy dispersive spectroscopy map of O concentration, which shows that these particles are likely copper oxides.**

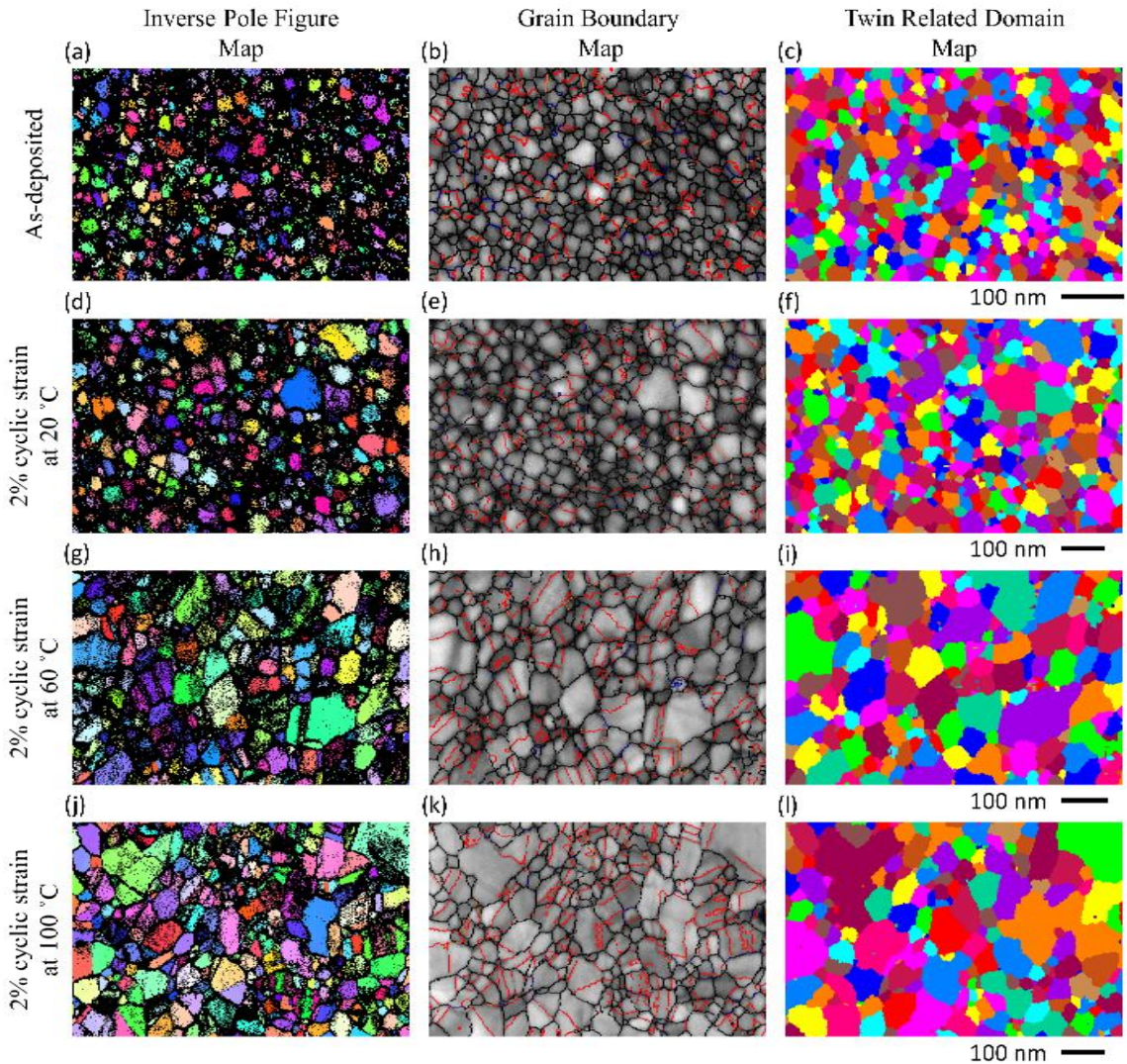

**Figure 4: The unprocessed inverse pole figures maps (IPFs), grain boundary maps, and twin related domain maps for each specimen, shown in the left, center, and right columns, respectively. The as-deposited material is shown in parts (a-c), 2% cyclic strain applied at 20 °C in (d-f), 2% cyclic strain applied at 60 °C in (g-i), and 2% cyclic strain applied at 100 °C in (j-l).**

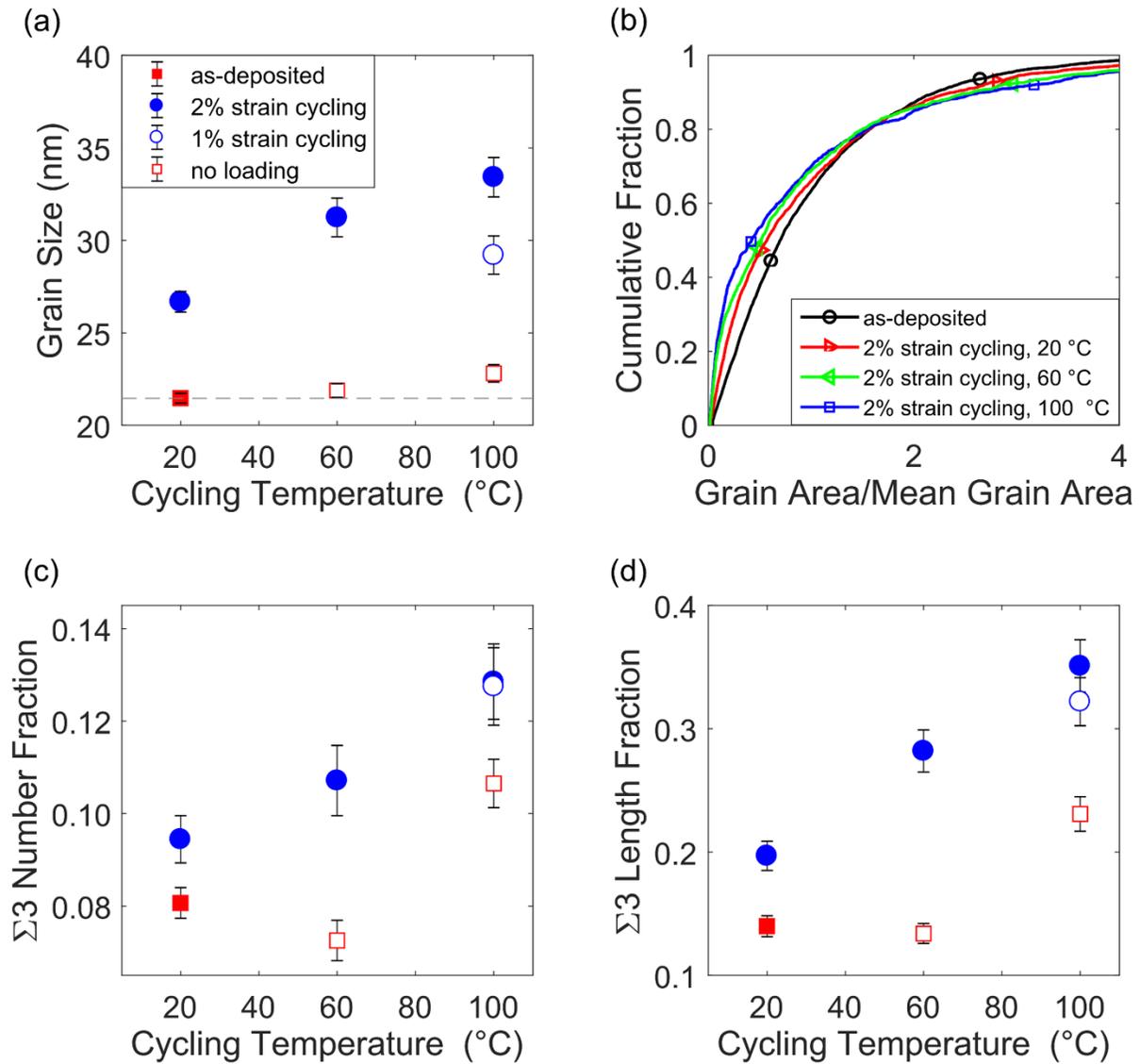

**Figure 5:** (a) The effect of cyclic strain and temperature on average grain size. (b) The normalized cumulative distribution functions for grain size of the as-deposited material and those exposed to 2% cyclic strain at several temperatures. (c) The Σ3 number fraction and (d) Σ3 length fraction of each material. Each metric increased noticeably with thermomechanical cycling. Parts (a), (c) and (d) share a common legend.

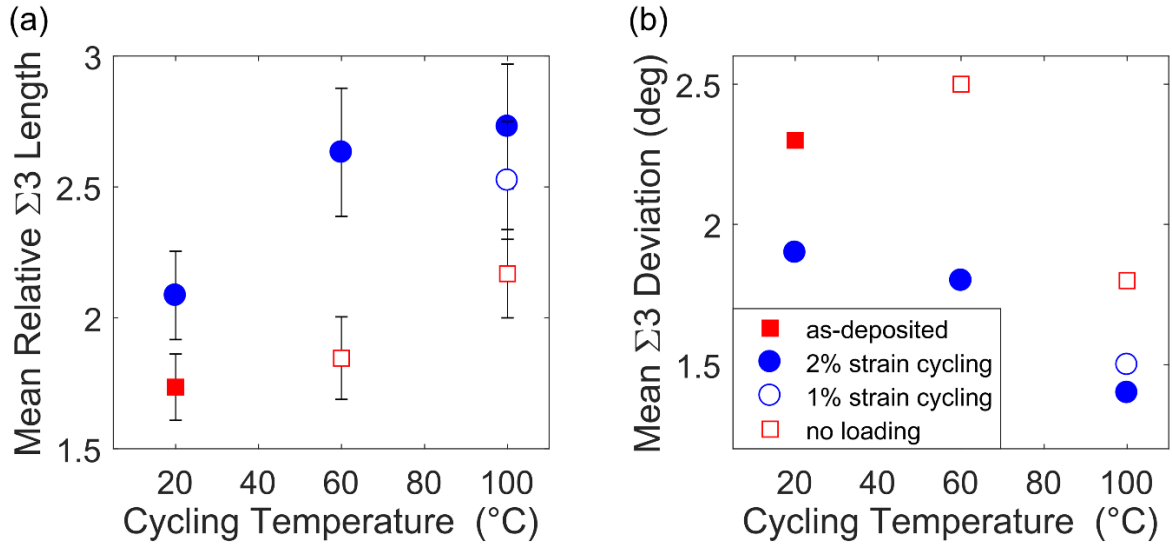

**Figure 6:** (a) The mean length of Σ3 boundaries relative to all other boundaries in each material. In every case, Σ3 boundaries were longer than average, with the amount increasing with thermomechanical cycling. (b) The mean deviation of the Σ3 boundaries from perfect CSL misorientation, which decreased with thermomechanical cycling.

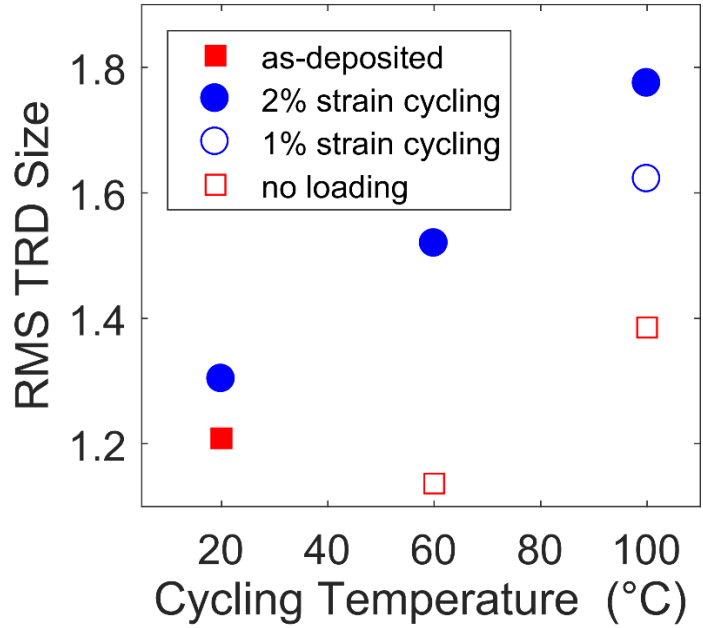

**Figure 7: The root mean squared (RMS) TRD size for each material. The increase in this parameter with cycling indicates that more grains became part of larger TRDs. Errors bars are smaller than symbols.**

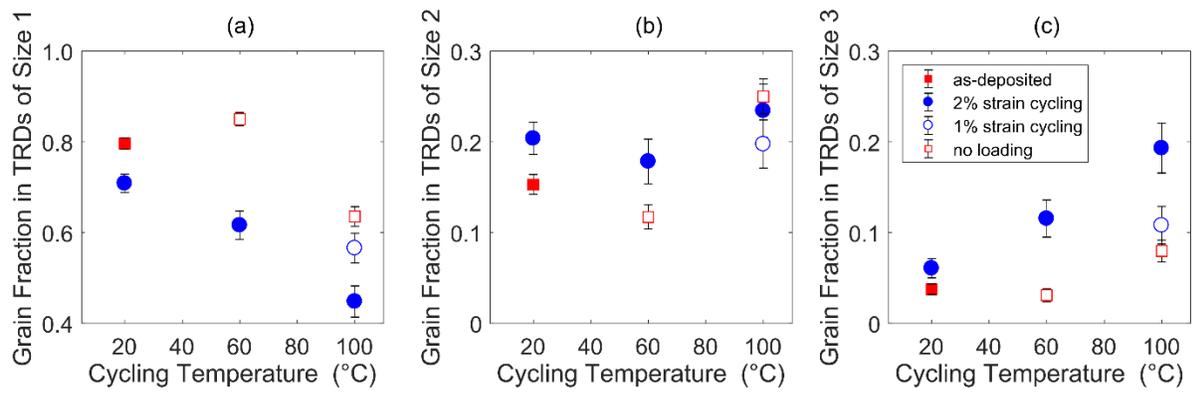
**Figure 8: The number fraction of grains in TRDs of (a) size 1, (b) size 2, and (c) size 3.**

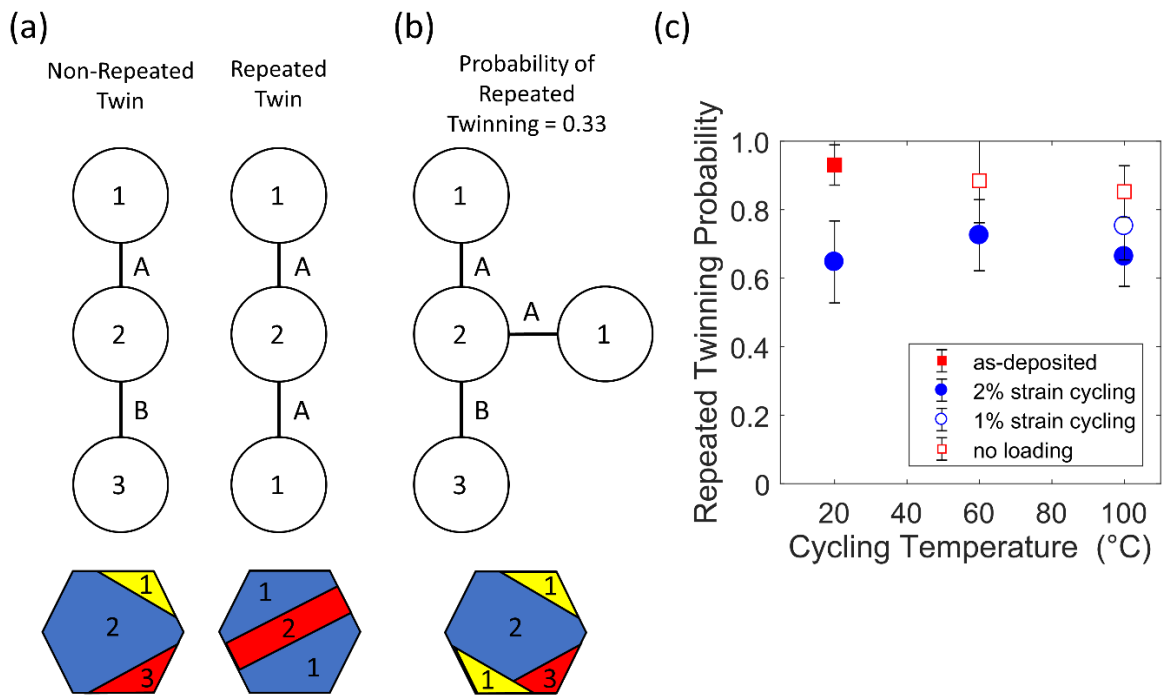

**Figure 9:** (a) Schematic of repeated versus non-repeated twinning. (b) Schematic of a 4-member TRD with a probability of repeated twinning of 0.33. In both (a) and (b), unique orientations are numbered and twin variants assigned letters 'A' or 'B.' (c) The probability of repeated twinning for 3-member TRDs.